%% file: main.tex
\documentclass[journal]{IEEEtran} 
\IEEEoverridecommandlockouts

\usepackage{booktabs}
\usepackage[utf8]{inputenc}
\usepackage{cite}
\usepackage{amsmath,amssymb,amsfonts}
\usepackage{algorithmic}
\usepackage{graphicx}
\usepackage{textcomp}
\usepackage{xcolor}
\usepackage{multirow}
\usepackage{verbatim}
\usepackage{stfloats}
\usepackage{tikz}

\usepackage[shortcuts,acronym]{glossaries}
\input{glossaries}

\def\BibTeX{{\rm B\kern-.05em{\sc i\kern-.025em b}\kern-.08em
    T\kern-.1667em\lower.7ex\hbox{E}\kern-.125emX}}

\newcommand\copyrighttext{%
  \footnotesize \textcopyright \the\year{} IEEE. Personal use of this material is permitted. Permission from IEEE must be obtained for all other uses, including reprinting/republishing this material for advertising or promotional purposes, collecting new collected works for resale or redistribution to servers or lists, or reuse of any copyrighted component of this work in other works.}

\newcommand\copyrightnotice{%
\begin{tikzpicture}[remember picture,overlay]
\node[anchor=north,yshift=0pt] at (current page.north) {\fbox{\parbox{\dimexpr0.9\textwidth-\fboxsep-\fboxrule\relax}{\copyrighttext}}};
\end{tikzpicture}%
}

\begin{document}
\bstctlcite{IEEEexample:BSTcontrol}

\title{Towards Trustworthy 6G Network Digital Twins: A Framework for Validating Counterfactual What-If Analysis in Edge Computing Resources}
 \author{
    \IEEEauthorblockN{
        Julian Jimenez Agudelo\IEEEauthorrefmark{1},
        Paola Soto \IEEEauthorrefmark{1},
        Ayat Zaki-Hindi\IEEEauthorrefmark{2},
        Jean-Sébastien Sottet\IEEEauthorrefmark{2},
        S\'ebastien Faye\IEEEauthorrefmark{2},\\
        Nina Slamnik-Kriještorac\IEEEauthorrefmark{1},
        Johann Marquez-Barja\IEEEauthorrefmark{1},
        Miguel Camelo Botero\IEEEauthorrefmark{1}}\\
    \IEEEauthorblockA{\IEEEauthorrefmark{1}University of Antwerp - imec, IDLab, Belgium}\\
    \IEEEauthorblockA{\IEEEauthorrefmark{2}Luxembourg Institute of Science and Technology (LIST), Luxembourg}
    \vspace*{-2em}
 }

\maketitle

\copyrightnotice 

\thispagestyle{empty}
\pagestyle{empty}


\begin{abstract}
\acp{ndt} enable safe what-if analysis for 6G cloud–edge infrastructures, but adoption is often limited by fragmented workflows from telemetry to validation. We present a data-driven \ac{ndt} framework that extends 6G-TWIN with a scalable pipeline for cloud–edge telemetry aggregation and semantic alignment into unified data models. Our contributions include: (i) scalable cloud–edge telemetry collection, (ii) regime-aware feature engineering capturing the network’s scaling behavior, and (iii) a validation methodology based on Sign Agreement and Directional Sensitivity. Evaluated on a Kubernetes-managed cluster, the framework extrapolates performance to unseen high-load regimes. Results show both \ac{dnn} and XGBoost achieve high regression accuracy ($R^2 > 0.99$), while the XGBoost model delivers superior directional reliability ($S_a > 0.90$), making the \ac{ndt} a trustworthy tool for proactive resource scaling in out-of-distribution scenarios.
\end{abstract}



\section{Introduction}

\acp{ndt} have emerged as a key enabler for future communication networks by providing a data-driven, interactive virtual replica of real network infrastructures. These twins support advanced emulation, scenario planning, impact analysis, and closed-loop optimization without directly affecting the physical network \cite{itu_t_y3090, ietf-nmrg-network-digital-twin-arch-07}. In particular, \acp{ndt} enable what-if analysis by offering predictive insights into the effects of configuration changes, traffic load variations, and resource allocation decisions, capabilities that are essential for operating the highly dynamic and distributed environments of 6G networks \cite{6g-twin-Vision, Ugent-ArchiDesign_for_NDT, RAZA2025108144}.

Despite this potential, the practical adoption of data-driven \acp{ndt} remains challenging. Existing research primarily focuses either on high-level architectural definitions and functional/data decompositions \cite{Zaki-Hindi2024, 6g-twin-framework, Zhiheng2024, itu_t_y3090} or on specialized models targeting narrow metrics with both real \cite{Saqib2024, zaki_hindi_2025_17086138} and synthetic data \cite{Polverini2026, FERRIOLGALMES2022109329, s21134321}. As a result, current approaches often overlook the \ac{e2e} operational workflows spanning data collection, preprocessing, and robust validation. This fragmentation limits the applicability of \acp{ndt} in real-world deployments where \acp{ndt} trustworthiness and robustness across varying operating regimes are critical.

Furthermore, data collection is frequently treated as an implicit process, despite being the foundation of any data-driven twin. In the cloud–edge continuum, heterogeneous data sources, in format and semantic, and dynamic service placements introduce noise and non-stationarity that directly impact model reliability. Therefore, ensuring a scalable and secured telemetry framework is a fundamental prerequisite for maintaining the semantic consistency and traceability of training data across distributed monitoring points \cite{6g-twin-framework, ietf-nmrg-network-digital-twin-arch-07}.

In this paper, we address these limitations by proposing an \ac{e2e} data-driven \ac{ndt} framework that bridges the gap between raw data collection and operational validation for computing environments. The primary contributions of this work are as follows:
\begin{itemize}
    \item We extend the 6G-TWIN architecture \cite{6g-twin-framework} with a unified telemetry and harmonization pipeline that decouples raw metric exposure from semantic abstraction, enabling consistent multi-domain observability across the cloud--edge continuum.
    \item We introduce a regime-aware feature engineering approach that abstracts raw infrastructure metrics into scale-invariant indicators, enabling the \ac{ndt} to extrapolate performance across heterogeneous workload intensities.
    \item We define a formal validation methodology for counterfactual reasoning based on Sign Agreement and Directional Sensitivity, providing a quantitative basis for assessing the trustworthiness of \ac{ndt} signals in autonomous \ac{mano} loops.
    \item Through an experimental validation on a Kubernetes-managed edge cluster, we show that while both \ac{dnn} and gradient-boosted models achieve high predictive accuracy, regime-aware XGBoost provides superior directional reliability under resource and workload transitions.
\end{itemize}

The remainder of the paper is organised as follows: Section \ref{sec:data-collection} details the data collection and harmonisation framework; Section \ref{sec:functional_modeling} presents the functional modelling pipeline and feature abstractions; Section \ref{sec:DT-Operation} describes NDT operation and matched-pairs validation; Section \ref{sec:results} reports the experimental results; Section \ref{conclusion} concludes.
\begin{figure}[t]
    \centering
\includegraphics[width=0.45\textwidth]{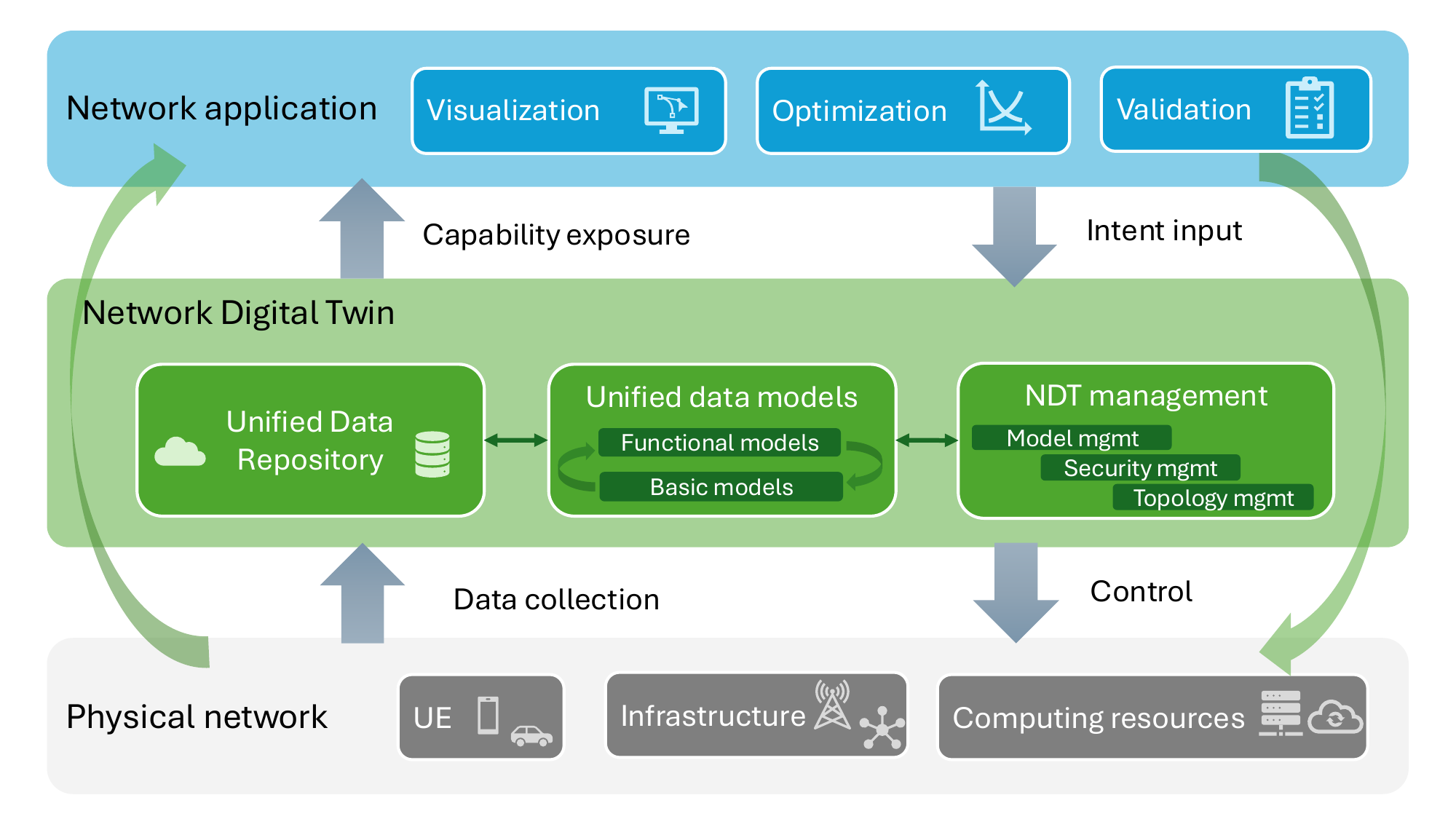}
    \caption{ITU Reference architecture \cite{itu_t_y3090}}
    \label{fig:Architectural_Components}
\end{figure}
\section{A Data Collection Framework for \ac{ndt}}\label{sec:data-collection}
This work builds upon the reference architecture proposed within the ITU \cite{itu_t_y3090}, which defines a modular framework for realizing \ac{ndt} through telemetry-driven data, model-based reasoning, and federated management.

\subsection{ITU Reference Architecture}
As illustrated in Fig.~\ref{fig:Architectural_Components}, the ITU-T high-level architecture enforces a clear separation of concerns across multiple layers: data acquisition and network control through interaction with the physical network via two different interfaces; real-world representation through the interaction of three key subsystems, an unified data and model repository and a management entity; an application layer that serves as the input for the requirements of the \ac{ndt}. These functions are coordinated by the \ac{mano} entity, enabling scalable and extensible deployment across heterogeneous network domains.

The lowest layer comprises the physical infrastructure, supporting various types of networks including the \ac{ran}, \ac{cn}, and \ac{ue}, supported by a device-to-edge-to-cloud continuum of computing resources. This layer provides potentially massive real-time operational data to the \ac{ndt} layer, together with an \textit{actuation and execution} interface that allows the \ac{ndt} to enforce optimized configurations on physical elements, thereby closing the control loop.

The \ac{ndt} layer represents the virtual counterpart of the physical network and is structured into two complementary functional views: (i) a \textbf{persistent view (real-time monitoring)}, which maintains a continuously updated representation of the current network state using a \ac{udr} and basic models; (ii) an \textbf{on-demand view (what-if analysis)}, which provides a sandbox environment for simulation and \ac{ai} training to assess hypothetical scenarios before real deployment. It relies on application-driven functional models, which may be either data- or simulation-based.


The \ac{mano} layer provides the \ac{ndt} lifecycle management by integrating \ac{zsm} principles. It orchestrates \ac{ai}-driven workflows and ensures that \ac{ndt} operations comply with high-level, intent-based policies, while also offering a unified interface for human operators to monitor network performance and intervene in automated decision-making. However, the current version of the ITU architecture does not explicitly define the functional structure of the data collection pipeline nor its interaction with other architectural components.

\begin{figure*}[t]
    \centering
    \includegraphics[width=0.65\textwidth]{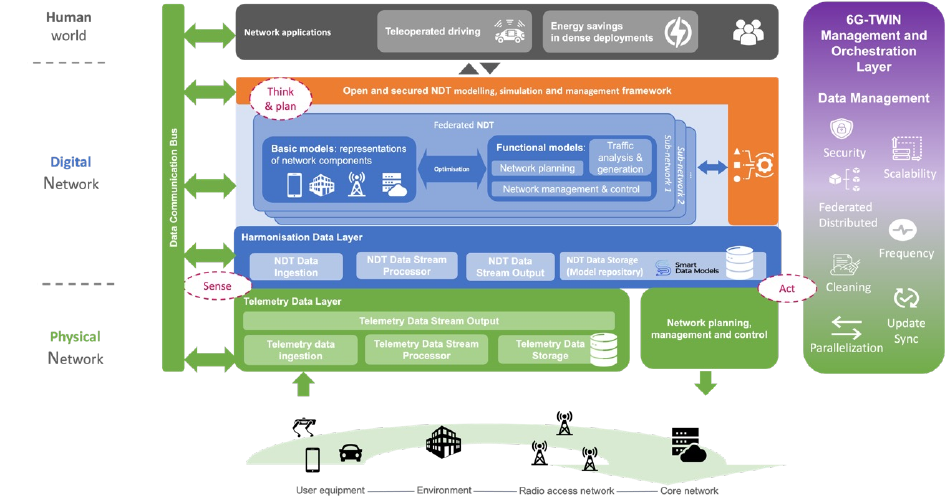}
    \caption{6G-TWIN enhanced architecture including the \ac{tdl} and \ac{hdl}}
    \label{fig:Architectural_data_collection}
\end{figure*}

\subsection{Data Collection and Harmonization in 6G-TWIN}


To address this gap, the 6G-TWIN data collection framework is extended as a multi-stage pipeline that bridges raw physical network observability with actionable \acf{ndt} models. This pipeline is explicitly partitioned into the \acf{tdl} and the \ac{hdl}, enabling tight synchronization across the cyber--physical continuum. 6G-TWIN defines an implementation-oriented functional architecture for 6G \acp{ndt} across the physical, digital, management, and application domains, refining the ITU-T vision into deployable functional blocks and interfaces \cite{6G-TWIN_D1.3_2025}.

This architectural extension is motivated by the limitations of existing generic telemetry frameworks when applied to 6G environments. While the IETF telemetry framework \cite{rfc9232} specifies core delivery and streaming mechanisms, it lacks domain-specific abstractions required for high-fidelity \ac{ndt} synchronization. Similarly, data-centric platforms such as \ac{ddsp} \cite{Padmanaban2024} provide scalable and real-time processing capabilities, but require adaptation to support telecommunications-specific data models and stringent latency constraints.

The new proposed \ac{tdl} and \ac{hdl} transform the traditional \textit{sense} phase into a specialized data pipeline capable of integrating heterogeneous operational data sources spanning physical and virtualized environments by separating raw telemetry ingestion from semantic refinement. As illustrated in Fig.~\ref{fig:Architectural_data_collection}, this dual-layered exposure framework replaces monolithic data collection approaches with a modular architecture optimized for efficient pre-processing, semantic alignment, and federated distribution.

\subsection{Multi-domain \acf{tdl}}

The multi-domain \ac{tdl} enables both real-time monitoring and historical analysis by collecting telemetry across management, control, and data planes, including the underlying computing infrastructure. Beyond periodic polling, modern telemetry supports \emph{query-driven and streaming} mechanisms, where measurement tasks are dynamically instantiated and exported to analytics components~\cite{rfc9232}. By leveraging structured data models and high-throughput transport, these approaches expose fine-grained system state with controlled overhead. Programmable telemetry architectures, such as~\cite{gupta2018sonata}, further unify configuration, performance, and flow-level observability, providing a scalable foundation for adaptive \acp{ndt}.


Extending beyond networking, the \ac{tdl} incorporates the \textbf{computing domain} to monitor cloud-native workloads hosted on \textbf{\ac{k8s}} clusters \cite{AI-DrivenDTF}. This is facilitated through integration with \textbf{Prometheus} \cite{SDN-Real-TimeMonitoring}, which acts as the primary time-series telemetry engine. By utilizing a pull-based scraping mechanism, Prometheus collects metrics from \ac{k8s} nodes, pods, and specialized exporters (e.g., \textit{cAdvisor}, \textit{Node Exporter}). These metrics, encompassing CPU/memory utilization, disk I/O, and microservice latency, are ingested into the \ac{tdl} and normalized. This cross-domain observability allows the \ac{ndt} to reason over the complex interdependencies between network throughput and the computational health of the host environment.



\subsection{\acf{hdl} for Semantic and Cross-Domain Data Models Harmonization}
The \ac{hdl} serves as the functional engine transforming normalized telemetry into \textbf{\ac{sdm}} \cite{smart_data_models}, bridging the gap between raw multi-domain data and the \textbf{\ac{udr}} \cite{Zaki-Hindi2024,sottet2025dataspace}. This layer addresses the heterogeneity of the 6G cyber-physical continuum by unifying 3GPP radio and core network models with the ETSI NFV-MANO framework for computing resources.

\begin{enumerate}
    \item \textbf{Data Ingestion and Taxonomy Mapping:} The \ac{hdl} standardizes naming conventions and data formats across diverse domains using a structured taxonomy of \textit{Attributes} and \textit{Performance Measurements}, also known as metamodels \cite{sottet2025dataspace}. For instance, radio assets follow 3GPP standards (e.g., TS 28.622 \cite{etsi_ts_128_622}). For the computing domain, it aligns with ETSI GS NFV-IFA standards to model NFV-IFA 011 and 013 \cite{etsi_ifa011, etsi_ifa013}:
    \begin{itemize}
        \item \textbf{(Virtual) \ac{ce}:} Modeling Virtual Machines and containers via resource allocations (CPU, memory, storage) and operational status.
        \item \textbf{\ac{nf} and \ac{ns}:} Representing individual functions (firewalls, routers) and their compositions (e.g., a VPN chain) as specialized \acp{ioc}.
        \item \textbf{Infrastructure \& Storage:} Mapping capacity and replication metrics from edge and cloud storage systems.
    \end{itemize}

\begin{figure}[t]
    \centering
    \includegraphics[width=0.47\textwidth]{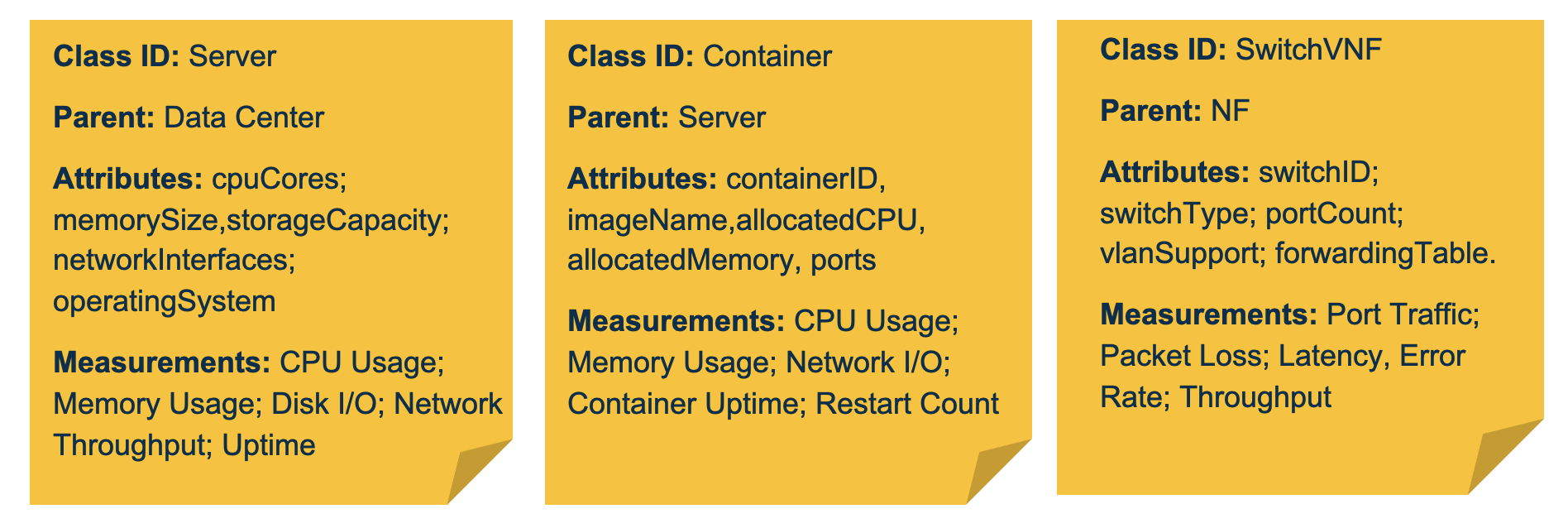}
    \caption{Example of metamodel schemas representing \acp{ce} in edge/cloud computing domains}
    \label{fig:ce_computing}
\end{figure}
    
    Notice that \ac{nf} and \ac{ns} may be described as a pure \acp{ce}, e.g., number of CPU, memory, storage, bandwidth, or from a more high-level perspective containing properties associated with its function or domain, e.g., an \ac{amf} or a router. An example of a schema to represent CE is presented in Fig. \ref{fig:ce_computing}.
    
    \item \textbf{Format Alignment:} To resolve discrepancies between heterogeneous vendors (e.g., RAN vs. Cloud Service Providers), the architecture adopts the \ac{sdm} approach.
    The proposed data model extends ETSI-defined descriptors. Beyond standardising the data format, it describes the extensive context in which data are provided, ensuring that virtualized infrastructure metrics (e.g., from \ac{k8s} or Prometheus) are  aligned with multi-domain network events. 

    \item \textbf{Semantic Alignment} The aforementioned \ac{sdm} definition ensures initial semantic consistency with respect to metric scales, units of measurement, and a standardized data format for subsequent exploitation. Building upon this foundation, semantic harmonization operates through complementary processes that enable consistent interpretation and comparability of measurements across diverse contexts and over time.  Context-aware adaptation mechanisms such as Chorus \cite{zhang2025chorus} help ensure that sensor outputs measuring the similar or complementary physical quantity remain interpretable and operationally consistent across varying environmental or situational contexts.

    \item \textbf{Basic Models Representation:} Once harmonized, the data is structured into basic models as a time-series graph capturing both temporal evolution and the relationships among network elements \cite{Zaki-Hindi2024}. This graph encodes physical and logical dependencies across components, providing a coherent and contextualized representation of the network state for visualization and for guiding subsequent processing and analysis by the functional models.
    
    \item \textbf{Functional Model Support:} Refined and harmonized data is fed into Functional Models for "what-if" analysis and predictive maintenance. In the computing domain, these models utilize harmonized data to forecast resource exhaustion, optimize \ac{vnf} placement, and detect anomalies in the compute-network chain to maintain end-to-end \ac{sla} requirements.
    
    \item \textbf{Data Space Governance:} In multi-stakeholder 6G environments, the \ac{hdl} prepares data for exchange within a \textbf{\ac{ds}}. This framework ensures data sovereignty for both 3GPP radio data and \ac{mano} logs (aligned with NFV-MANO policies), enabling secure collaboration between \acp{mno} and cloud providers.
\end{enumerate}

Ultimately, the 6G-TWIN data collection framework serves as a unified semantic substrate, formally converging 3GPP Generic Network Resource Models (\textit{via} ETSI TS 128 622) and ETSI NFV-MANO specifications into a cohesive \ac{sdm} ecosystem. This integration harmonizes the multi-domain \ac{tdl} in the physical world with the \ac{hdl} in the digital world, achieving a multidimensional abstraction of \acp{nf} and \acp{ns} across the edge-to-cloud continuum. These entities are represented simultaneously as deterministic \acp{ce} (resource-level footprints) and high-level functional \acp{ioc}.  

This dual-lens representation, embedded within a sovereign \ac{ds} governance framework, enables the \ac{ndt} to reason over complex interdependencies between radio performance and the computational health of the virtualized infrastructure, a prerequisite for autonomous 6G orchestration.

\section{Constructing Functional Models for Computing-Centric \acp{ndt}}
\label{sec:functional_modeling}
\begin{figure}[t]
    \centering
    \includegraphics[width=0.5\textwidth]{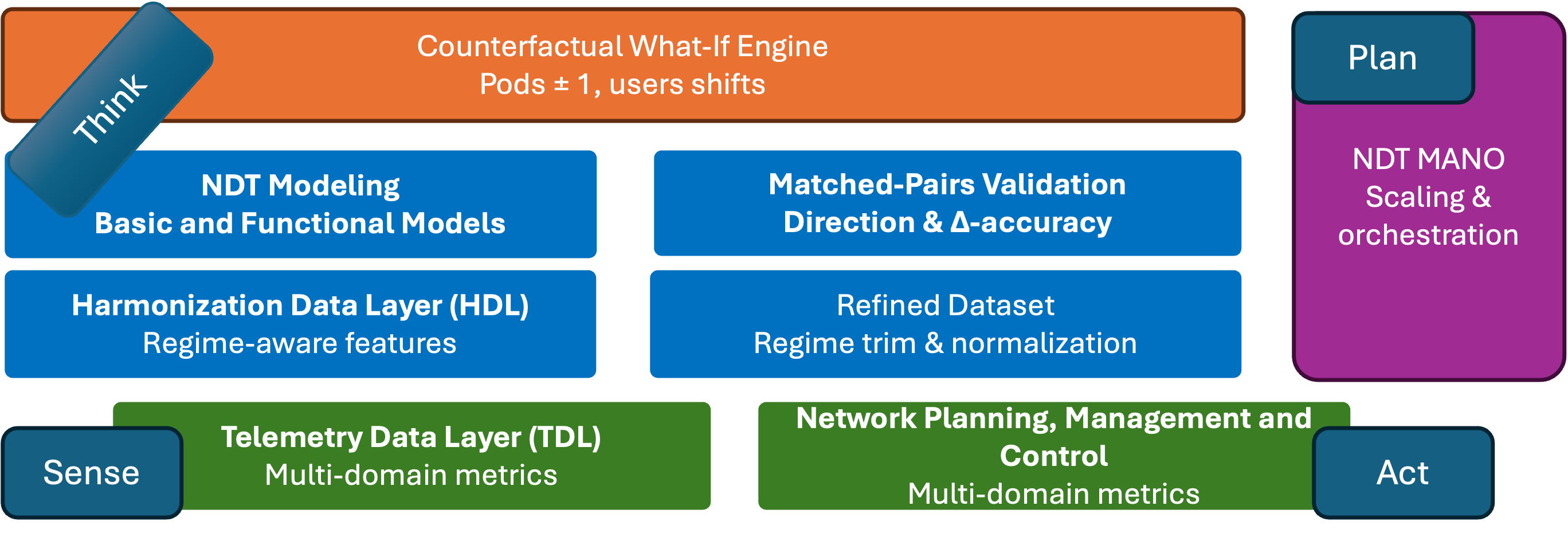}
    \caption{NDT Framework for Validating Counterfactual What-If Analysis in Edge Computing Resources}
    \label{fig:NDT framework}
\end{figure}

Building upon the previously introduced \ac{tdl} and \ac{hdl}, this section presents the framework for constructing functional models in computing-centric \acp{ndt}. Although the Sense--Think--Plan--Act loop is native to the 6G-TWIN architecture, we instantiate it here for the edge computing use case by tightly coupling the collection layers with functional modeling blocks for counterfactual what-if analysis (cf.~Fig.~\ref{fig:NDT framework}). This specialization preserves generality, as the modeling and validation components remain domain-agnostic and extensible.

In the \emph{Sense} phase, multi-domain telemetry gathered and normalized by the \ac{tdl} is harmonized by the \ac{hdl} into regime-aware features. These representations feed the \emph{Think} phase, where basic and functional NDT models capture structural and dynamic aspects of cloud--edge infrastructures, and a counterfactual engine evaluates controlled perturbations (e.g., pod scaling $\pm 1$, user shifts) to simulate alternative states.

Trustworthiness is ensured through matched-pairs validation assessing directionality and $\Delta$-accuracy between predicted and observed outcomes, producing a calibrated dataset. The resulting insights drive the \emph{Plan} and \emph{Act} phases via NDT-MANO, enabling informed scaling and orchestration decisions while keeping the modeling layer logically decoupled from runtime control.


\subsection{Modeling Hierarchy: Structural and Behavioral Abstractions}
The modeling process follows a hierarchical approach that separates static system representation from dynamic behavior modeling.
\textbf{Basic Models} provide a continuously synchronized structural representation of the physical and virtual infrastructure. Aligned with ETSI GS NFV-IFA specifications, these models describe:
(i) \emph{Element-level properties}, such as deterministic resource footprints of \acp{ce} (CPU, memory, storage, and bandwidth), and functional attributes of \acp{nf}; and
(ii) \emph{System-level topologies}, capturing placement constraints, resource sharing, and service chaining relationships. These models form the digital mirror of the infrastructure and are maintained through real-time updates stored in the \ac{udr}.

\textbf{Functional Models} extend this structural baseline by learning the relationship between resource allocation, workload dynamics, and service-level performance. These models constitute the predictive core of the \ac{ndt}, enabling forecasting of latency, saturation effects, and scaling behavior under varying conditions. Unlike basic models, functional models encode non-linear dependencies across the compute--network stack and are explicitly designed to support what-if reasoning.

\subsection{Feature Construction and Regime-Aware Abstraction}\label{sec:Feat_Construction_Regime-Aware}

To ensure model generalization across heterogeneous operating conditions, we employ a structured methodology for \textbf{regime-aware feature abstraction}. This approach moves beyond the ingestion of raw telemetry by deriving indicators that capture the underlying scaling physics and resource dynamics of the distributed environment.

The pipeline is divided into two primary phases: statistical refinement and physics-informed feature engineering.

\subsubsection{Regime-Specific Statistical Refinement}
Prior to feature extraction, raw data is partitioned into discrete \textit{operational regimes} defined by the intersection of workload intensity (e.g., user concurrency) and resource allocation (e.g., active pod counts). Within each regime, an automated anomaly removal process is executed using the Interquartile Range (IQR) method. By filtering outliers on a per-regime basis, the pipeline ensures that the models learn from steady-state system behavior rather than transient measurement artifacts or "cold-start" noise, which can otherwise skew the learned scaling relationships.

\subsubsection{Physics-Informed Feature Engineering}
Rather than relying on absolute system counters, the pipeline constructs features that represent the "pressure" and "flow" of the infrastructure. This abstraction strategy includes:

\begin{itemize}
    \item \textbf{Resource Utilization Density:} Metrics such as CPU and memory usage are processed to represent the load per unit of compute. This allows the model to learn load-to-performance relationships that are invariant to specific cluster sizes.
    \item \textbf{Congestion and Queue Dynamics:} Indicators derived from the average queue depth and backlog time provide a direct signal of system saturation. These serve as proxy indicators for latent bottlenecks that raw throughput metrics often fail to capture.
    \item \textbf{Scale-Invariant Transformations:} By centering the feature space on the relationship between demand (workload) and capacity (allocated resources), the framework enables \textit{out-of-distribution (OOD) generalization}. This allows the Digital Twin to maintain predictive validity when extrapolated to higher-intensity traffic regimes (e.g., 600 users) despite being trained on lower-intensity data (e.g., 200--400 users).
\end{itemize}

Through this abstraction layer, the resulting functional models become portable across different deployment sizes and hardware configurations. Additionally, the models provide reliable "What-If" projections for proactive infrastructure scaling and capacity planning by preserving the physical consistency of system behavior.

\subsection{Model Outputs and Decision Interfaces}

The functional models, acting as the core of the \ac{ndt}, expose predictive outputs through standardized interfaces designed for proactive system orchestration. Rather than static performance monitoring, this layer provides dynamic mapping from system state to expected \ac{qos} metrics.

\subsubsection{Predictive Estimations}
The primary output is a point estimate of the target latency, denoted as $y = f(\mathbf{x}, \mathbf{c})$, where $\mathbf{x}$ represents the current telemetry state and $\mathbf{c}$ denotes a specific infrastructure configuration (e.g., pod count). The models are optimized to provide high-fidelity inference even at the distribution tails, ensuring that the orchestrator can account for worst-case latency scenarios in high-concurrency regimes.

\subsubsection{Counterfactual (What-If) Analysis}
To support automated decision-making, the \ac{ndt} facilitates \textit{counterfactual reasoning}. By perturbing the configuration vector $\mathbf{c} \to \mathbf{c}'$ (e.g., simulating a $\pm 1$ pod adjustment), the model generates relative performance trends. As evidenced in our experimental validation, these "What-If" projections are verified against ground-truth matched pairs from the dataset. This allows the orchestrator to quantify the expected latency delta ($\Delta y$) before an actuation command is issued to the live environment.

\subsubsection{Operational Decoupling}
Critically, the functional layer remains agnostic to specific deployment policies or optimization objectives (e.g., cost vs. performance). By providing a pure performance estimation interface, also know as \textbf{Analyzing NDT} \cite{ETSI_ZSM_015}, the model preserves a separation of concerns: the \ac{ndt} models the system physics, while the operational methodology, described in the following section, utilizes these predictions to derive optimal scaling trajectories.

\section{Methodology for NDT Operation and Trustworthy Evaluation}
\label{sec:DT-Operation}
The operational lifecycle of a 6G-TWIN analytical instance governs how structural and functional models are instantiated, trained, validated, and used for decision support. This methodology ensures that predictive outputs generated by the \ac{ndt} remain trustworthy when applied to operational what-if analysis and closed-loop actuation.

\subsection{Initialization and Structural Synchronization}
Upon instantiation, the \ac{ndt} performs a consistency check on the Basic Model repository. If no valid synchronized state exists, the 6G-TWIN \ac{mano} triggers an infrastructure discovery procedure via the \ac{tdl}. This process initiates the capture and normalization of new telemetry data, instantiating two distinct data flows. This decoupled approach ensures that data is simultaneously available for live structural mirroring, enabling real-time updates to the basic models, and persistent on-demand analysis, providing the historical telemetry required to train or refine functional models.

Depending on the specific analytical objectives (e.g., the required time-window of observation), normalized telemetry is ingested and processed by the \ac{hdl} and mapped into \acp{ioc} representing computing and networking entities, such as \ac{k8s} nodes, pods, and service chains. These entities are then bound to live telemetry streams, establishing a high-fidelity structural mirror. This "digital twin" of the infrastructure serves as the authoritative reference state for all subsequent modeling, what-if simulations, and closed-loop actuation.

\subsection{Functional Model Training and Dataset Assembly}
Functional model training is initiated on demand by the \ac{ndt} \ac{mano}. A bounded and normalized telemetry window is extracted from the \ac{udr} and processed to generate training datasets aligned with specific operating regimes. Feature construction follows the regime-aware abstractions defined in Section~\ref{sec:functional_modeling}, ensuring that learned relationships reflect system physics rather than transient correlations.

Multiple model classes (e.g., tree-based and neural architectures) may be trained in parallel, enabling comparative evaluation without committing to a single hypothesis.

\subsection{Trustworthy Evaluation via Matched-Pairs Consistency}
To assess suitability for operational what-if analysis, the framework introduces a \textbf{Matched-Pairs Consistency} evaluation methodology. Rather than relying solely on aggregate prediction errors, this approach evaluates model behavior across controlled transitions between paired operating regimes $\mathcal{R}_A \rightarrow \mathcal{R}_B$, where a single control variable is modified.

Two complementary metrics are computed:
\begin{enumerate}
    \item \textbf{Sign Agreement ($S_a$):} the probability that the model correctly predicts the direction of change in the target \ac{kpi}.
    \item \textbf{Directional Sensitivity:} the consistency between predicted and observed performance deltas across regime pairs.
\end{enumerate}

Based on these metrics, models are assigned a \emph{Qualitative Deployment Grade} (Excellent, Reliable, or Unreliable). This grade acts as a deployment gate, ensuring that only models exhibiting stable directional behavior under perturbations are promoted to the active \ac{ndt} runtime.

\subsection{What-If Execution and Actuation}
Validated models enable interactive what-if exploration. Once the (best) functional model is selected and deployed, the operators can start simulating hypothetical scaling actions or workload changes, which are evaluated by the functional model to predict performance trends and resource risks. When confidence thresholds are met, the \ac{ndt} \ac{mano} can translate validated decisions into actuation commands, such as adjusting the \ac{k8s} scheduler, thereby closing the loop between digital reasoning and physical execution.

\section{Experimental Evaluation and Results}
\label{sec:results}

\subsection{System Implementation and Experimental Campaign}

The proposed framework is validated on a three-node \ac{k8s} cluster (Table~\ref{tab:system_specs}) using \textbf{Prefect} for orchestration and \textbf{KEDA} for workload-driven scaling. Telemetry ingestion and harmonization are realized through a unified data pipeline built on \textbf{Prometheus} and \textbf{Kafka}, which jointly serve as a communication bus and distributed streaming substrate. Prometheus collects raw infrastructure and service-level metrics, while Kafka enables durable and decoupled data dissemination across the cloud--edge continuum. Lightweight processing components perform online feature extraction and semantic alignment, effectively instantiating both the \ac{tdl} and \ac{hdl} while preserving separation of concerns. 


The telemetry is persisted in \textbf{InfluxDB}, providing a historical basis for \ac{hdl}-driven model lifecycle management. During training phases, the \ac{hdl} processes these data streams to generate functional models (\ac{dnn} and XGBoost), which are subsequently versioned and stored in a \textbf{container registry}. Associated performance metadata (e.g., MAE, $R^2$, and regime validity) are stored to ensure reproducible and traceable orchestration decisions.

\textit{Data Generation Strategy:} A structured measurement campaign is conducted to characterize non-linear system behavior under load. A synthetic load generator issues continuous inference requests to a \texttt{CIFAR-10} classification service across three workload regimes, 200, 400, and 600 concurrent users. For each workload, the service is scaled from 1 to 6 pods, covering operating points from saturation to over-provisioning. Telemetry is collected via the \ac{tdl} at 20\,ms resolution, yielding 259,016 samples. This granularity is required to capture queuing dynamics and tail latency behavior under contention.

\begin{table}[t]
\centering
\caption{System Infrastructure and Software Environment}
\label{tab:system_specs}
\renewcommand{\arraystretch}{1.2}
\resizebox{\columnwidth}{!}{%
\begin{tabular}{@{}lccc@{}}
\toprule
\textbf{Attribute} & \textbf{Master Node} & \textbf{Worker 1} & \textbf{Worker 2} \\ \midrule
CPU Cores & 4 (AMD Ryzen 9) & 8 (Intel i5-7260U) & 8 (Intel i5-7260U) \\
RAM (GB) & 8 & 8 & 32 \\
OS / K8s & \multicolumn{3}{c}{Ubuntu 22.04 / Kubernetes v1.33.6} \\ \midrule
Component & Version & Component & Version \\ \midrule
Prefect & 2.20.18 & KEDA & 2.18.1 \\
InfluxDB & 2.x & Kafka & 3.x \\ \bottomrule
\end{tabular}%
}
\end{table}

\subsection{Data Refinement and Regime-Aware Abstraction}

The dataset consists of telemetry samples collected via the \ac{tdl}, which performs real-time acquisition and persistence into InfluxDB and exposes streams through Kafka for online consumption. For functional model construction, the \ac{hdl} retrieves these measurements and applies a structured refinement pipeline.

Anomaly removal is applied independently for each $(\text{users}, \text{pods})$ operating regime using percentile-based trimming at the 1st and 99th percentiles of the target latency metric \texttt{avg\_total\_infer\_ms}. This removes transient cold-start effects and measurement artifacts, resulting in a refined dataset of 253,828 samples.

The refined telemetry is mapped into a 7-dimensional regime-aware feature space designed to capture non-linear scaling dynamics. As detailed in Table~\ref{tab:feature_transformation}, raw metrics are normalized by the number of active pods to derive scale-invariant indicators such as workload intensity and congestion index. This abstraction allows the models to recognize equivalent operating states, for example 400 users on 4 pods and 200 users on 2 pods. To stabilize variance induced by heavy-tailed latency distributions, the target variable is log-transformed as $y' = \ln(1 + y)$.

To evaluate out-of-distribution generalization, models are trained exclusively on the 200 and 400 user regimes and evaluated on the unseen 600 user regime.

\begin{table}[t]
\centering
\caption{Regime-Aware Feature Engineering}
\label{tab:feature_transformation}
\renewcommand{\arraystretch}{1.3}
\resizebox{\columnwidth}{!}{%
\begin{tabular}{@{}lll@{}}
\toprule
\textbf{Raw Telemetry} & \textbf{Derived Feature} & \textbf{Physical Meaning} \\ \midrule
\texttt{current\_users}, \texttt{pods} & Workload Intensity & Demand per processing unit \\
\texttt{avg\_depth\_on\_enqueue}, \texttt{pods} & Congestion Index & Queue pressure per pod \\
\texttt{avg\_backlog\_sec\_est}, \texttt{pods} & Backlog Flow & Delay accumulation rate \\
\texttt{avg\_cpu\_process\_pct}, \texttt{avg\_cpu\_system\_pct} & Compute Pressure & Aggregate CPU load \\
\texttt{avg\_mem\_system\_pct} & Memory Density & Secondary bottleneck signal \\
\texttt{avg\_total\_infer\_ms} & Log-Target Scaling & Tail variance stabilization \\ \bottomrule
\end{tabular}%
}
\end{table}

\subsection{Model Configuration and Performance Evaluation}

Two functional model architectures are evaluated within the \ac{ndt}, an XGBoost regressor and a \ac{dnn}, both trained on the regime-aware feature space to predict log-transformed inference latency.

\subsubsection{Architectural Configuration}

The XGBoost model is configured with 2,000 estimators, a maximum depth of 8, and a learning rate of 0.05. The \ac{dnn} consists of three fully connected layers with 256, 128, and 64 neurons, respectively, and incorporates batch normalization, dropout rates between 0.3 and 0.6, and a Huber loss optimized using Adam with a learning rate of $10^{-3}$.

\subsubsection{NDT Quality Analysis}

Model quality is evaluated on the unseen 600 user regime using the criteria defined in Table~\ref{tab:quality_criteria}. As summarized in Table~\ref{tab:model_performance}, both models achieve an \textit{Excellent} quality grade. While overall accuracy is comparable, XGBoost exhibits superior robustness in high-latency regions.

\begin{table}[t]
\centering
\caption{NDT Quality Grading Criteria}
\label{tab:quality_criteria}
\renewcommand{\arraystretch}{1}
\resizebox{\columnwidth}{!}{%
\begin{tabular}{@{}llcc@{}}
\toprule
Quality Grade & Constraint Type & MAE Bound & $P_{95}$ Bound \\ \midrule
Excellent & Absolute & $\le 50$ ms & $\le 150$ ms \\
          & Relative & $\le 10\%$ & $\le 25\%$ \\ \addlinespace
Good      & Absolute & $\le 100$ ms & $\le 300$ ms \\
          & Relative & $\le 15\%$ & $\le 40\%$ \\ \addlinespace
Weak      & Combined & \multicolumn{2}{c}{Criteria not satisfied} \\ \bottomrule
\end{tabular}%
}
\end{table}

\begin{figure*}[t]
\centering
\begin{tabular}{cc}

\includegraphics[width=0.45\textwidth]{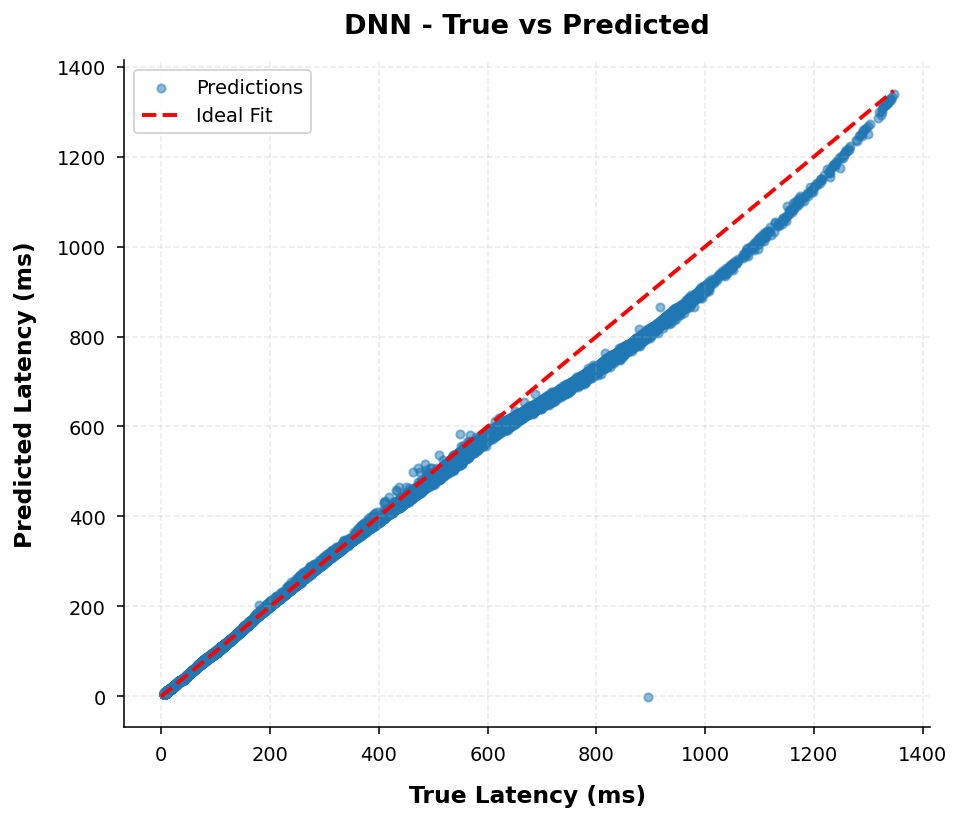} &
\includegraphics[width=0.45\textwidth]{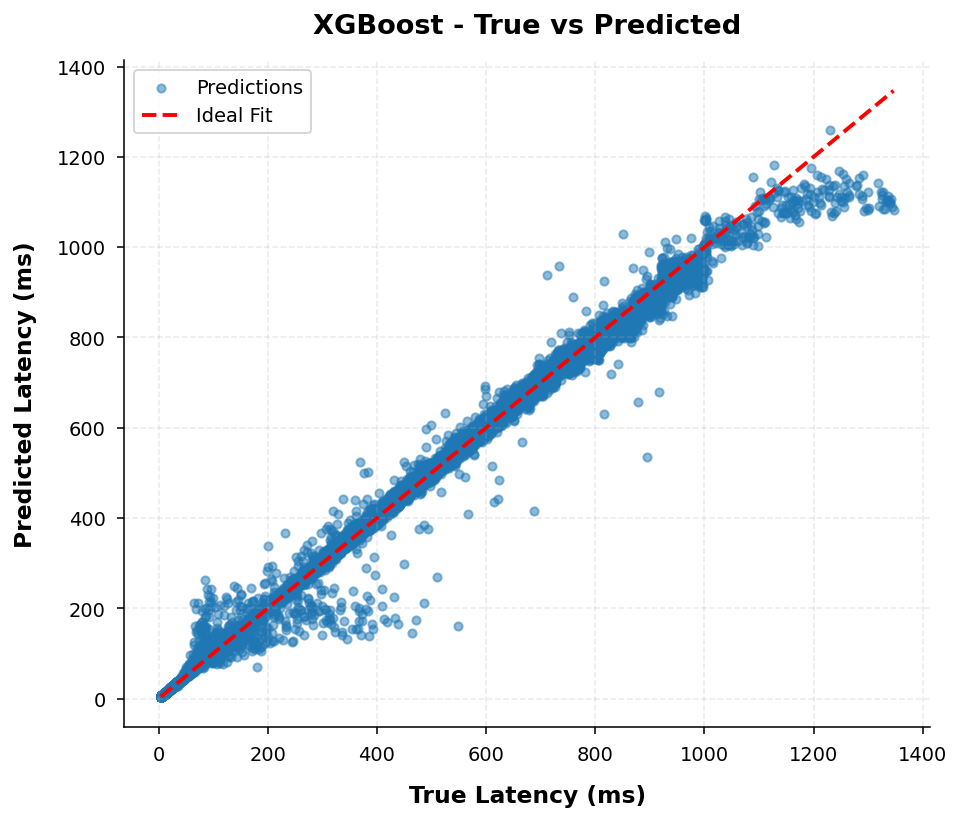} \\
\end{tabular}
\caption{True versus predicted latency for the unseen 600 user regime. The dashed diagonal indicates ideal prediction.}
\label{fig:pred_composite}
\end{figure*}

\begin{table}[t]
\centering
\caption{NDT Performance Summary (600 User Test Regime)}
\label{tab:model_performance}
\renewcommand{\arraystretch}{1.3}
\resizebox{\columnwidth}{!}{%
\begin{tabular}{lccccc}
\toprule
Model & MAE (ms) & RMSE (ms) & $R^2$ & P95 Error (ms) & Grade \\ \midrule
XGBoost & 8.97 & 18.08 & 0.9949 & 29.32 & Excellent \\
DNN     & 9.03 & 18.69 & 0.9945 & 50.35 & Excellent \\ \bottomrule
\end{tabular}%
}
\end{table}

\begin{table}[th]
\centering
\caption{Tail Performance Analysis (600 User Test Regime)}
\label{tab:tail_performance}
\renewcommand{\arraystretch}{1.3}
\resizebox{\columnwidth}{!}{%
\begin{tabular}{llcccc}
\toprule
Threshold & Model & Samples & MAE (ms) & RMSE (ms) & $R^2$ \\ \midrule
$P_{90}$ & XGBoost & 4,555 & 23.00 & 37.85 & 0.8952 \\
         & \ac{dnn}     & 4,555 & 51.28 & 54.49 & 0.7829 \\ \addlinespace
$P_{95}$ & XGBoost & 2,278 & 30.43 & 48.53 & 0.8049 \\
         & \ac{dnn}     & 2,278 & 61.80 & 65.10 & 0.6490 \\ \bottomrule
\end{tabular}%
}
\end{table}

\subsubsection{Per-Regime Validation}

Prediction accuracy remains stable across pod allocations within the 600 user regime, as shown in Table~\ref{tab:regime_accuracy}. XGBoost maintains particularly high fidelity for 5 and 6 pod configurations.

\begin{table}[t]
\centering
\caption{Per-Regime Accuracy (600 User Test)}
\label{tab:regime_accuracy}
\renewcommand{\arraystretch}{1.3}
\resizebox{\columnwidth}{!}{%
\begin{tabular}{@{}llcccc@{}}
\toprule
Users & Pods & XGBoost MAE & XGBoost $R^2$ & \ac{dnn} MAE & \ac{dnn} $R^2$ \\ \midrule
600 & 4 & 11.26 & 0.9881 & 8.96 & 0.9959 \\
600 & 5 & 8.56  & 0.9956 & 9.23 & 0.9951 \\
600 & 6 & 8.83  & 0.9955 & 9.02 & 0.9943 \\ \bottomrule
\end{tabular}%
}
\end{table}

\subsection{Operational What-If Analysis}

To evaluate the NDT for decision-oriented network and service management, we perform what-if reasoning by estimating latency deltas $\Delta$ between observed operational regimes and hypothetical configuration or workload changes. Each transition is defined using matched ground-truth samples with similar nuisance conditions, ensuring consistent and physically meaningful delta computations.

As shown in Table~\ref{tab:what_if_analysis}, the XGBoost-based NDT captures both direction and magnitude of performance changes. For horizontal scaling at 600 users, mean $\Delta = -12.85$\,ms for pods $+1$ and $-0.01$\,ms for pods $-1$. For workload reductions, mean $\Delta = -12.24$\,ms ($600\rightarrow400$) and $-20.50$\,ms ($600\rightarrow200$). Median deltas and $P_{95} |\Delta|$ show strong alignment with observed changes and high sensitivity to system dynamics, confirming suitability for proactive network management and resource orchestration.

The \ac{dnn} shows limited response to counterfactual changes, with mean $\Delta = 0.60$\,ms (pods $+1$) and $-0.92$\,ms (pods $-1$), and small deltas for workload reductions ($-0.09$\,ms for $600\rightarrow400$, $0.45$\,ms for $600\rightarrow200$). While adequate for static latency prediction, it lacks directional sensitivity required for reliable service management decisions under large operational shifts.

In general, we can see that XGBoost model provides consistent directional predictions and accurately quantifies the impact of resource and workload changes, supporting closed-loop orchestration and operational what-if reasoning. The \ac{dnn} remains limited to static predictions and cannot reliably capture causal effects of configuration or load modifications.

\begin{table}[t]
\centering
\caption{Counterfactual Delta and Directional Sensitivity Analysis}
\label{tab:what_if_analysis}
\renewcommand{\arraystretch}{1.2}
\resizebox{\columnwidth}{!}{%
\begin{tabular}{@{}llcccl@{}}
\toprule
\textbf{Scenario} & \textbf{Model} & \textbf{Mean $\Delta$ (ms)} & \textbf{$S_a$} & \textbf{MAE$_{\Delta}$ (ms)} & \textbf{Sensitivity} \\ \midrule
\textit{pods +1}  & \ac{dnn}     & 0.60    & 0.52 & 4.12 & Negligible \\
                  & XGBoost & -12.85  & 0.94 & 2.15 & High \\ \addlinespace
\textit{pods -1}  & \ac{dnn}     & -0.92   & 0.48 & 6.84 & Negligible \\
                  & XGBoost & 24.85   & 0.91 & 3.42 & High \\ \addlinespace
\textit{users $\to$ 200} & \ac{dnn} & 0.45 & 0.55 & 5.11 & Low \\
                  & XGBoost & -20.50  & 0.98 & 1.88 & High \\ \bottomrule
\end{tabular}%
}
\end{table}


\section{Conclusions and Future Work}\label{conclusion}

This paper presented a telemetry driven approach to functional modeling within the 6G-TWIN reference architecture, focusing on the evaluation of data driven models for \ac{e2e} latency prediction in a \ac{ndt}. The results showed that non-linear models such as \ac{xgboost} and \ac{dnn} can achieve high predictive accuracy, but may exhibit different biases and sensitivities that are not captured by aggregate error metrics alone. In particular, the evaluation under what-if scenarios demonstrated that models with similar pointwise accuracy can differ in their ability to estimate the impact of configuration and load changes, highlighting the need for evaluation criteria that reflect operational reliability rather than purely statistical performance.

Future work will extend the proposed evaluation framework by incorporating additional classes of functional models and more complex what-if scenarios, including time dependent and multi dimensional variations. Further investigation will also focus on improving model robustness and reducing bias in high latency regimes, as well as on integrating evaluation outcomes more tightly with network management functions. These directions aim to strengthen the role of data driven digital twins as reliable decision support tools in future communication networks.


\section*{Acknowledgment}
The 6G-TWIN project has received funding from the Smart Networks and Services Joint Undertaking (SNS JU) under the European Union’s Horizon Europe research and innovation program under Grant Agreement No 101136314.


\bibliographystyle{IEEEtran}
\bibliography{sources.bib}

\end{document}

%% file: glossaries.tex
\newacronym{vni}{VNI}{Visual Networking Index}
\newacronym{ism}{ISM}{Industrial, Scientific, and Medical}
\newacronym{5g}{5G}{Fifth Generation}
\newacronym{6g}{6G}{Sixth Generation}
\newacronym{b5g}{B5G}{Beyond 5G}
\newacronym{m2m}{M2M}{Machine-to-Machine}
\newacronym{pid}{PID}{Proportional–Integral–Derivative}
\newacronym{pi}{PI}{Proportional–Integral}
\newacronym{cpu}{CPU}{Central Processing Unit}
\newacronym{ram}{RAM}{Random Access Memory}
\newacronym{sla}{SLA}{Service Level Agreement}
\newacronym{slo}{SLO}{Service Level Objective}
\newacronym{fifo}{FIFO}{First In, First Out}
\newacronym{Sim-Diasca}{Sim-Diasca}{Simulation of Discrete Systems of All Scales}
\newacronym{thd}{THD}{Threshold}
\newacronym{dc}{DC}{Data Center}
\newacronym{e2e}{E2E}{End-to-End}
\newacronym{6g-twin}{6G-TWIN}{6G-TWIN}
\newacronym{ce}{CE}{Computing Element}
\newacronym{ne}{NE}{Network Element}

\newacronym{ran}{RAN}{Radio Access Network}
\newacronym{upf}{UPF}{User Plane Function}
\newacronym{vran}{vRAN}{Virtualized Radio Access Network}
\newacronym{cn}{CN}{Core Network}
\newacronym{tn}{TN}{Transport Network}
\newacronym{ns}{NS}{Network Services}
\newacronym{vn}{VN}{Virtual Network}
\newacronym{vnf}{VNF}{Virtual Network Function}
\newacronym{pnf}{PNF}{Physical Network Function}
\newacronym{cnf}{CNF}{Cloud-Native Network Function}
\newacronym{vm}{VM}{Virtual Machine}
\newacronym{sdn}{SDN}{Software Defined Networking}
\newacronym{nfv}{NFV}{Network Function Virtualization}
\newacronym{mano}{MANO}{Management and Orchestration}
\newacronym{mec}{MEC}{Multi-access Edge Computing}
\newacronym{sfc}{SFC}{Service Function Chaining}
\newacronym{ue}{UE}{User Equipment}
\newacronym{wifi}{Wi-Fi}{IEEE 802.11}
\newacronym{wlan}{WLAN}{Wireless Local Area Network}
\newacronym{ap}{AP}{Access Point}
\newacronym{sta}{STA}{Station}
\newacronym{csma}{CSMA/CA}{Carrier-Sense Multiple Access with Collision Avoidance}
\newacronym{mcs}{MCS}{Modulation and Coding Scheme}
\newacronym{csa}{CSA}{Channel Switch Announcement}
\newacronym{cb}{CB}{Channel Bonding}
\newacronym{dcb}{DCB}{Dynamic Channel Bonding}
\newacronym{scb}{SCB}{Static Channel Bonding}
\newacronym{rssi}{RSSI}{Received Signal Strength Indicator}
\newacronym{sinr}{SINR}{Signal-to-Interference plus Noise Ratio}
\newacronym{obss}{OBSS}{Overlapping Basic Service Set}
\newacronym{bss}{BSS}{Basic Service Set}
\newacronym{sr}{SR}{Spatial Reuse}
\newacronym{dcf}{DCF}{Distributed Coordination Function}
\newacronym{ofdma}{OFDMA}{Orthogonal Frequency Division Multiple Access}
\newacronym{cca}{CCA}{Channel Clear Assessment}
\newacronym{ed}{ED}{Energy Detection}

\newacronym{lte}{LTE}{Long-Term Evolution}
\newacronym{lte-u}{LTE-U}{LTE Unlicensed}
\newacronym{lte-laa}{LTE-LAA}{LTE Licensed Assisted Access}
\newacronym[longplural={Transmission Opportunities}]{txop}{TXOP}{Transmission Opportunity}

\newacronym{sdr}{SDR}{Software Defined Radio}
\newacronym{usrp}{USRP}{Universal Software Radio Peripheral}
\newacronym{cr}{CR}{Cognitive Radio}
\newacronym{imt}{IMT}{International Mobile Telecommunications}
\newacronym{cu}{CU}{Centralized Unit}
\newacronym{du}{DU}{Distributed Unit}
\newacronym{nh}{NH}{Neutral Host}
\newacronym{nhn}{NHN}{Neutral Host Network}
\newacronym{mno}{MNO}{Mobile Network Operator}

\newacronym{ai}{AI}{Artificial Intelligence}
\newacronym{ml}{ML}{Machine Learning}
\newacronym{dl}{DL}{Deep Learning}
\newacronym{lr}{LR}{Linear Regression}
\newacronym{dtr}{DTR}{Decision Tree Regression}
\newacronym{dnn}{DNN}{Deep Neural Network}
\newacronym{xgboost}{XGBoost}{eXtreme Gradient Boosting}

\newacronym{rl}{RL}{Reinforcement Learning}
\newacronym{drl}{DRL}{Deep Reinforcement Learning}
\newacronym{dqn}{DQN}{Deep Q-Network}
\newacronym{pomdp}{POMDP}{Partially Observable Markov Decision Process}
\newacronym{mdp}{MDP}{Markov Decision Process}
\newacronym{ddpg}{DDPG}{Deep Deterministic Policy Gradients}
\newacronym{ppo}{PPO}{Proximal Policy Optimization}
\newacronym{a2c}{A2C}{Advantage Actor Critic}
\newacronym{trpo}{TRPO}{Trust Region Policy Optimization}
\newacronym{aco}{ACO}{Ant Colony Optimization}
\newacronym{pso}{PSO}{Particle Swarm Optimization}
\newacronym{mlops}{MLOps}{Machine Learning Operations}

\newacronym{sl}{SL}{Supervised Learning}
\newacronym{cnn}{CNN}{Convolutional Neural Network}
\newacronym{fnn}{FNN}{Feedforward Neural Network}
\newacronym{nn}{NN}{Neural Network}
\newacronym{mlp}{MLP}{Multilayer Perceptron}
\newacronym{ae}{AE}{Autoencoder}
\newacronym{relu}{ReLU}{Rectified Linear Unit}
\newacronym{gb}{GB}{Gradient Boosting}
\newacronym{svm}{SVM}{Support Vector Machine}
\newacronym{rf}{RF}{Random Forest}
\newacronym{dt}{DT}{Decision Tree}
\newacronym{lstm}{LSTM}{Long Short-Term Memory}

\newacronym{gnn}{GNN}{Graph Neural Network}
\newacronym{gcn}{GCN}{Graph Convolutional Network}
\newacronym{gcl}{GCL}{Graph Convolutional Layer}
\newacronym{gnb}{gNB}{gNodeB}
\newacronym{sb3}{SB3}{Stable-Baselines3}

\newacronym{ndt}{NDT}{Network Digital Twin}
\newacronym{tndt}{TNDT}{Telemetry-driven Network Digital Twin}
\newacronym{tdl}{TDL}{Telemetry Data Layer}
\newacronym{hdl}{HDL}{Harmonization Data Layer}
\newacronym{dt-c}{DT-C}{Digital Twin Connector}
\newacronym{amqp}{AMQP}{Advanced Message Queuing Protocol}
\newacronym{dds}{DDS}{Data Distribution Service}
\newacronym{mqtt}{MQTT}{Message Queuing Telemetry Transport}
\newacronym{coap}{CoAP}{Constrained Application Protocol}
\newacronym{nats}{NATS}{Neural Autonomic Transport System}
\newacronym{udr}{UDR}{Unified Data Repository}

\newacronym{qos}{QoS}{Quality of Service}
\newacronym{qoe}{QoE}{Quality of Experience}
\newacronym{rmse}{RMSE}{Root Mean-Squared Error}
\newacronym{kpi}{KPI}{Key Performance Indicator}
\newacronym{wfq}{WFQ}{Weighted Fair Queuing}
\newacronym{drr}{DRR}{Deficit Round Robin}
\newacronym{red}{RED}{Random Early Detection}

\newacronym{itu}{ITU}{International Telecommunication Union}
\newacronym{etsi}{ETSI}{European Telecommunications Standards Institute}

\newacronym{zsm}{ZSM}{Zero-Touch Network and Service Management}
\newacronym{cli}{CLI}{Command Line Interface}
\newacronym{snmp}{SNMP}{Simple Network Management Protocol}
\newacronym{nf}{NF}{Network Function}
\newacronym{k8s}{K8s}{Kubernetes}
\newacronym{mae}{MAE}{Mean Absolute Error}
\newacronym{bmp}{BMP}{BGP Monitoring Protocol}
\newacronym{int}{INT}{In-band Network Telemetry}
\newacronym{ds}{DS}{Data Space}
\newacronym{sdm}{SDM}{Smart Data Models}
\newacronym{ioc}{IOC}{Information Object Classes}
\newacronym{ddsp}{DDSP}{Distributed Data Stream Processing}
\newacronym{amf}{AMF}{Access and Mobility Management Function}